\documentclass[twocolumn,superscriptaddress,aps,preprintnumbers,amsmath,amssymb,prl,nofootinbib,compress
]{revtex4-1}

\usepackage{amsfonts}
\usepackage{ascmac}
\usepackage{bm}
\usepackage[caption=false]{subfig}
\usepackage{comment}
\usepackage{ifpdf}
\usepackage{slashed}
\usepackage{color}
\usepackage[mathscr]{eucal}
\usepackage[utopia]{mathdesign}

\usepackage{cancel}

\ifpdf
  \usepackage{graphicx}     
  \usepackage[bookmarksopen,colorlinks=true,linkcolor=bblue,citecolor=ppink,urlcolor=ppink]{hyperref}
\else     
  \usepackage[dvipdfmx]{graphicx}     
  \usepackage[dvipdfmx,bookmarksopen,colorlinks=true,linkcolor=bblue,citecolor=ppink,urlcolor=ppink]{hyperref}
\fi

\definecolor{red}{rgb}{1,0,0}
\definecolor{ppink}{rgb}{1,0.4,0.4}
\definecolor{bblue}{rgb}{0.400000,0.400000,1.000000}

\newcommand{\prn}[1]{\left( {#1} \right)}

\begin{document}


\title{
Electroweak Vacuum Stabilized by Moduli during/after Inflation
}

\author{Yohei Ema}
\affiliation{Department of Physics, Faculty of Science, The University of Tokyo}
\author{Kyohei Mukaida}
\affiliation{Kavli IPMU (WPI), UTIAS, University of Tokyo, Kashiwa, 277-8583, Japan}
\author{Kazunori Nakayama}
\affiliation{Department of Physics, Faculty of Science, The University of Tokyo}
\affiliation{Kavli IPMU (WPI), UTIAS, University of Tokyo, Kashiwa, 277-8583, Japan}

\begin{abstract}
\noindent
It is known that the present electroweak vacuum is likely to be 
metastable and it may lead to a serious instability during/after inflation.
We propose a simple solution to the problem of vacuum instability during/after inflation.
If there is a moduli field which has Planck-suppressed interactions with the standard model fields,
the Higgs quartic coupling in the early universe naturally takes a different value from the present one.
A slight change of the quartic coupling in the early universe makes the Higgs potential absolutely stable
and hence we are free from the vacuum instability during/after inflation.
\end{abstract}

\date{\today}
\maketitle
\preprint{UT 16-22}
\preprint{IPMU 16-0075}

\section{Introduction}
\label{sec:}
\setcounter{equation}{0}

After the discovery of the 125\,GeV Higgs boson~\cite{Aad:2012tfa,Chatrchyan:2012xdj}, 
one of the interesting cosmological issues is the stability of the electroweak vacuum.
If we take the center value of the measured top quark mass~\cite{ATLAS:2014wva}, 
the quartic coupling becomes negative at energy scale of $\sim 10^{10}\,$GeV, 
and hence the present electroweak vacuum is metastable~\cite{Arnold:1989cb,Sher:1988mj,Anderson:1990aa,Arnold:1991cv,Espinosa:1995se,Isidori:2001bm,Espinosa:2007qp,Ellis:2009tp,Bezrukov:2009db,Degrassi:2012ry,Buttazzo:2013uya}.\footnote{
	It may be possible to make the electroweak vacuum absolutely stable 
	by, \textit{e.g.} introducing
	an additional Higgs portal singlet scalar which acquires a large vacuum expectation value.
	See Refs.~\cite{Lebedev:2012zw,EliasMiro:2012ay}.
} Although the lifetime of the present vacuum is much longer than the age of the universe,
it is non-trivial whether or not
the Higgs falls into unwanted deeper minimum in the early universe.

First let us suppose that there is no large effective mass term for the Higgs during inflation.
During inflation, the infrared (IR) Higgs fluctuations develop.
The typical amplitude of the quantum fluctuation generated during the one Hubble time is $\sim \mathcal H_{\rm inf}/2\pi$ with $\mathcal H_{\rm inf}$ being the Hubble scale during inflation.
This process can be viewed as a classical random walk process~\cite{Starobinsky:1994bd}.
As a result, after $N$ e-foldings, the mean Higgs field value acquires
\begin{align}
	\sqrt{\left< h^2 \right>}
	 \simeq \sqrt N \frac{\mathcal H_{\rm inf}}{2\pi},
\end{align}
where $h$ is the field value of the physical Higgs boson.
The classical motion overcomes this quantum noise for $h\gtrsim h_c$ where
\begin{align}
	h_c^2 = \frac{3\mathcal H_{\rm inf}^4}{8\pi^2 m^2_{\rm eff}(h_c)} ~~~\leftrightarrow~~~ 
	h_c\simeq \frac{0.4 \mathcal H_{\rm inf}}{\lambda^\frac{1}{4}}.
	\label{hc}
\end{align}
Here $m_{\rm eff}^2(h)=\lambda h^2$ is the effective mass of the Higgs.
This means that the natural value of the Higgs field value during inflation is 
\begin{align}
	h \simeq {\rm min}\left[ \sqrt N \frac{\mathcal H_{\rm inf}}{2\pi} ,~ h_c\right].
\end{align}
Since the total e-folding number $N$ must be larger than $\sim 50$, we can reasonably take $h\sim h_c$.

So far we have assumed that the Higgs quartic coupling $\lambda$ is positive independently of the Higgs field value.
However, it is actually indicated that $\lambda$ becomes negative 
at high energy scale (which we denote by $h_{\rm max}$) 
due to the loop effect caused by the large top yukawa coupling.\footnote{
	Later we will define $h_{\rm max}$ in a slightly different manner, but practically the precise definition is not important.
}
It is clear that it leads to a disaster if $h_{\rm max} < h_c$: in this case the Higgs falls into the true vacuum during inflation
and the present electroweak vacuum is never realized thereafter~\cite{Lebedev:2012sy,Fairbairn:2014zia,Hook:2014uia,Herranen:2014cua,Kamada:2014ufa,Kearney:2015vba,Espinosa:2015qea}.

The vacuum instability during inflation is easily avoided by introducing a Higgs-inflaton and/or Higgs-curvature coupling like
\begin{align}
	V =  cI^2 |H|^2 ~~{\rm and/or}~~c' R |H|^2,   \label{IH}
\end{align}
where $I$ and $R$ are the inflaton field and Ricci scalar, respectively and
$H$ denotes the Higgs doublet. 
These couplings generate large mass terms for the Higgs field and hence the development of IR fluctuations can be suppressed.
Even in this case, however, we must take care of the vacuum instability occurring {\it after} inflation,
since these additional mass terms rapidly oscillate during the inflaton oscillation era
and it leads to the resonant enhancement of the Higgs fluctuations.
To avoid the catastrophe, upper bounds on these coupling constants are obtained~\cite{Herranen:2015ima,Ema:2016kpf,Kohri:2016wof}.
Combined with the requirement of the vacuum stability during inflation,
there is only a small window for the parameter region of these coupling constants.\footnote{
	In addition, even if the resonant Higgs production 
	does not cause
	the decay of the electroweak vacuum during the preheating stage,
	it could happen \textit{afterwards}. This is because the cosmic expansion 
	reduces the effective mass term induced by 
	the Higgs-inflaton and/or Higgs-curvature coupling
	so that the Higgs fluctuations may eventually overcome the potential barrier.
	Thermalized population of other SM particles might  save this situation,
	but it strongly depends on thermalization processes and further studies are required.
}

In this letter we propose a different approach for the issue of the vacuum stability during/after inflation.
The crucial observation is that the Higgs quartic coupling in the early universe needs not coincide with that of the present value.
In particular, it may depend on the value of some scalar field, $\phi$, which we call moduli.
Since the field value of moduli during/after inflation can be different from the present one,
it is natural to expect that the quartic coupling in the early universe is also different from the present one.
If the additional contribution to the quartic coupling is the same order as the present one,
the absolute stability of the Higgs potential during/after inflation is ensured.

\section{Electroweak vacuum stabilized by moduli}
\label{sec:vac}

We consider a moduli field $\phi$ which has Planck-suppressed interactions with standard model (SM) fields.
For the stability of Higgs, the most important coupling is the moduli coupling to the Higgs:
\begin{align}
	V_h =\left( \lambda + c_\lambda\frac{\phi-v_\phi}{M_P}\right) |H|^4,
\end{align}
where $c_\lambda$ is the coupling constant of $\mathcal O(1)$, $v_\phi$ is the present vacuum expectation value (VEV) of moduli and $M_P$ is the reduced Planck scale.
The potential of the moduli is assumed to be
\begin{align}
	V_\phi = \frac{1}{2}m_\phi^2(\phi-v_\phi)^2 + \frac{C_H^2}{2}\mathcal {H}^2 \phi^2,
\end{align}
where $\mathcal H$ denotes the Hubble parameter, $m_\phi$ is the moduli mass and $C_H$ is a coupling constant.\footnote{
	Without loss of generality, we can shift the moduli field such that the Hubble mass term takes the form of $\sim \mathcal H^2\phi^2$
	and the coupling constants coincide with the present values at the potential minimum $\phi=v_\phi$.
	We take this convention. 
}
It ensures that in the early universe $\mathcal H \gtrsim m_\phi$, the moduli sits at $\phi \simeq m_\phi^2v_\phi/(m_\phi^2+C_H^2 \mathcal H^2)$.
When the moduli is displaced from the minimum $v_\phi$, the effective quartic coupling is given by
\begin{align}
	\lambda(\phi) = \lambda_0 + c_\lambda\frac{\phi-v_\phi}{M_P}.
	\label{eq:lam_phi}
\end{align}
Here and in what follows the subscript $0$ indicates that the quantity is evaluated at the present vacuum $\phi=v_\phi$.
In particular, during/after inflation at which $C_H\mathcal H \gg m_\phi$ (and hence $|\phi| \ll v_\phi$), it is approximately given by
\begin{align}
	\lambda(\phi) \simeq \lambda_0-\frac{c_\lambda v_\phi}{M_P} \equiv \lambda_0 -\xi_\lambda.    \label{xilam}
\end{align}
Since $\xi_\lambda$ is naturally expected to be $\mathcal O(0.1\mathchar`-\mathchar`-1)$, 
it significantly modifies the Higgs potential in the early universe and even the vacuum $H=0$ can be absolutely stable.\footnote{
	We regard the potential as ``absolutely stable'' if the potential remains positive up to $h \simeq M_P$.
	See also Refs.~\cite{Branchina:2013jra,Branchina:2014rva} for possible effects of 
	higher-dimensional Planck-suppressed operators on the vacuum stability.
}

Not only the quartic coupling, but also the top yukawa coupling $y_t$ is modified 
if the moduli has a coupling like
\begin{align}
	&\mathcal L =   \left( y_{t0} + c_y\frac{\phi-v_\phi}{M_P} \right) \overline Q_t \tilde H t_R + {\rm h.c.},
	\label{eq:yuk_phi}
\end{align}
where $y_{t0}$ is the present top yukawa coupling, $Q_t$ is the left-handed top quark doublet, $t_R$ is the right-handed top quark and $c_y$ is a coupling constant of $\mathcal O(1)$.
The effective top yukawa coupling is given by
\begin{align}
	y_t (\phi)= y_{t0} + c_y\frac{\phi-v_\phi}{M_P}.
\end{align}
Similarly to the quartic coupling, for $C_H\mathcal H \gg m_\phi$ it is approximately given by
\begin{align}
	y_t(\phi) \simeq y_{t0} -\frac{c_y v_\phi}{M_P} \equiv y_{t0} -\xi_y.   \label{xiy}
\end{align}
It is known that the Higgs potential is very sensitive to the top mass (or top yukawa) and
even a few percent decrease of the top mass compared with the center value makes the Higgs potential absolutely 
stable~(See \textit{e.g.}, \cite{Degrassi:2012ry}).
Since $\xi_y$ is naturally expected to be $\mathcal O(0.1\mathchar`-\mathchar`-1)$, 
it can also significantly modify the Higgs potential in the early universe through the radiative correction.\footnote{
	Flavor symmetry may suppress the coupling of the moduli to other SM quarks and leptons
	so that they are less important than the top yukawa coupling for the vacuum stability.
}

The gauge coupling constants can also be modified in a similar fashion by introducing the moduli couplings like
\begin{align}
	\mathcal L =  -\left( 1+ c_{g_i}\frac{\phi-v_\phi}{M_P} \right)\frac{1}{4} F_{\mu\nu}^a F^{\mu\nu a}.
\end{align}
For $\mathcal H \gg m_\phi$, the gauge couplings $g_i$, with $i=1,2,3$ corresponding to the U(1), SU(2) and SU(3) gauge groups, become
\begin{align}
	\frac{1}{g_i^2(\phi)} \simeq \frac{1}{g_{i0}^2}  \left(1-\frac{c_{g_i} v_\phi}{M_P}\right) \equiv \frac{1}{g_{i0}^2}(1-\xi_{g_i}).
\end{align}
All of these modifications of coupling constants by the moduli significantly affect the stability of electroweak vacuum
in the early universe.
For simplicity, below we consider non-zero $\xi_\lambda$ and $\xi_y$ only.\footnote{
	The moduli may also couple to the Higgs kinetic term.
	However, such a coupling is translated into Eqs.~\eqref{eq:lam_phi}
	and~\eqref{eq:yuk_phi} after canonically normalizing the Higgs for constant $\phi$.
	Hence we neglect it here.
}

\begin{figure}[t]
\begin{center}
\begin{tabular}{cc}
\includegraphics[width=\linewidth]{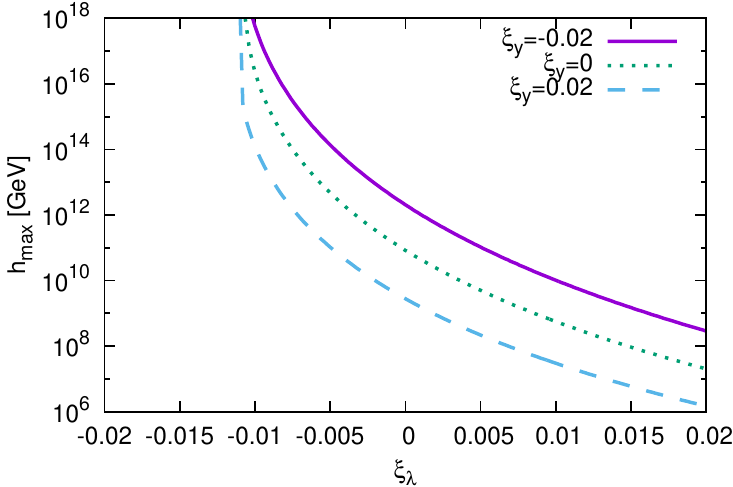}
\end{tabular}
\end{center}
\caption {\small
$h_{\rm max}$ as a function of $\xi_\lambda$ for several choices of $\xi_y$.
The instability scale $h_{\rm max}$ blows up at $-\xi_\lambda \sim \mathcal{O}(0.01)$ because
$\lambda(\phi=0)$ at the Planck scale becomes positive at around this point.
}
\label{fig:hmax}
\end{figure}

We have calculated the effective potential of the Higgs at the one-loop order
for non-zero $\xi_\lambda$ and $\xi_y$ according to Ref.~\cite{Degrassi:2012ry},
\begin{align}
	V_h = \frac{\lambda_{\rm eff}(h)}{4}h^4.
\end{align}
We have imposed the boundary condition (\ref{xilam}) and (\ref{xiy}) at the Planck scale
and define $h_{\rm max}$ by $\left(\partial V_h / \partial h\right)_{h=h_{\rm max}}= 0$.
We have chosen $\lambda_0$ and $y_{t0}$ at the Planck scale such that
they reproduce the current central values of the Higgs and top quark masses 
at the electroweak scale when $\xi_\lambda = \xi_y = 0$.
Fig.~\ref{fig:hmax} plots $h_{\rm max}$ as a function of $\xi_\lambda$ for several choices of $\xi_y$.
It is seen that $-\xi_\lambda \sim \mathcal O(0.01)$ is sufficient to ensure the absolute stability of the Higgs.
Note that we may need at least
$h_{\rm max} \gtrsim \mathcal H_{\rm inf}$ 
for the vacuum stability because there may be a infrared cutoff 
at the energy scale of $\sim \mathcal H_{\rm inf}$ during inflation.

Here we comment on some subtleties related to the effective potential.
It is known that an effective potential is generally 
gauge dependent~\cite{Jackiw:1974cv}.
Although extrema of the effective potential are 
formally gauge independent~\cite{Nielsen:1975fs,Fukuda:1975di}, 
still some care is needed to maintain the gauge independence of the extrema 
in the perturbative calculation~\cite{Andreassen:2014gha}.
One should take care of these subtleties for the precise calculation of,
\textit{e.g.}, the life time of the electroweak vacuum~\cite{Bednyakov:2015sca}.
Our main purpose here is, however, not to precisely determine
$\xi_\lambda$ and $\xi_y$ needed for the absolute stability of the Higgs,
but to demonstrate our key idea.
Hence our treatment is enough for that purpose.

\section{After inflation}
\label{sec:after}

In the previous section we have seen that the presence 
of the moduli can make the Higgs potential absolutely stable
as long as the moduli $\phi$ is displaced from its potential minimum.
However, the deeper minimum appears after the moduli starts to oscillate.
In order for the Higgs not to fall into the deeper minimum, the amplitude of the Higgs condensate or the typical fluctuation of the 
Higgs must be smaller than $\sim 10^{10}\,$GeV at $\mathcal H \sim m_\phi$.
This leads to a constraint on the moduli mass.
Let us now see the dynamics of the Higgs and moduli after inflation.

\subsection{Before moduli oscillation}

Here we briefly study the dynamics of Higgs condensate after inflation, 
but before the moduli starts to oscillate: $m_\phi \lesssim \mathcal H < \mathcal H_\text{inf}$.

First, let us suppose that the Higgs condensate develops as (\ref{hc}) during inflation.
Just after inflation, it starts to oscillate and it behaves as relativistic matter since the potential is quartic.
At the early stage of Higgs oscillation, the resonant production of weak gauge bosons happens~\cite{Figueroa:2015rqa,Enqvist:2015sua}
while the Higgs amplitude decreases due to the Hubble expansion.
The inflaton decay also produces high-temperature plasma which scatters off the Higgs condensate.
It acts as the dissipation effect on 
the Higgs condensate~\cite{Berera:1995ie,Berera:2008ar,BasteroGil:2009ec,BasteroGil:2010pb,Yokoyama:2004pf,Yokoyama:2005dv,Drewes:2010pf,Drewes:2013iaa,Mukaida:2012qn,Mukaida:2012bz,Mukaida:2013xxa}.
Both processes tend to thermalize the Higgs condensate.

If, on the other hand, the Higgs is strongly stabilized at the origin due to the coupling like (\ref{IH})
during inflation, highly inhomogeneous Higgs fluctuations develop 
after the inflaton starts to oscillate~\cite{Herranen:2015ima,Ema:2016kpf,Kohri:2016wof}. 
The efficiency of weak gauge boson production is somehow reduced at the first stage
compared with the case of homogeneous Higgs condensate.
Also in this case, the dissipation effect of the plasma produced by the inflaton decay
makes the system close to thermalized plasma.

In order to investigate the dissipation effect,
we first estimate the time when the SM particles produced via the inflaton decay are thermalized.
To be concrete, suppose that the inflaton reheats the universe via Planck-suppressed operators.
In this case, the thermalization rate is estimated as
$\Gamma_\text{th} \sim \alpha^2 T^\text{(w.b.)} \sqrt{T^\text{(w.b.)}/m_I}$,
where $T^\text{(w.b.)}$ indicates  ``would-be'' temperature of radiation when it is thermalized,
$m_I$ represents the inflaton mass,
and $\alpha = g^2/(4\pi)$, 
with $g$ collectively denoting the gauge and yukawa couplings.
Comparing it with the Hubble parameter,
one obtains the thermalization time~\cite{Harigaya:2013vwa,Mukaida:2015ria}
\begin{align}
	t_\text{th}^{-1} 	\sim \alpha^\frac{16}{5} m_I  \prn{ \frac{T_\text{R}^2 M_P}{m_I^3} }^\frac{3}{5}
\end{align}
where $T_{\rm R}$ denotes the reheating temperature.
The thermalization temperature of radiation is then estimated to be~\cite{Harigaya:2013vwa,Mukaida:2015ria}
\begin{align}
	T_{\rm th} 
	&\sim \alpha^{4/5} m_I \prn{\frac{T_{\rm R}^2 M_P}{m_I^3}}^\frac{2}{5} \nonumber \\
	&\sim 9 \times 10^{11}\, \text{GeV}\,
	\prn{ \frac{\alpha}{0.1} }^\frac{4}{5} \prn{ \frac{10^{13}\, \text{GeV}}{m_I} }^\frac{1}{5} 
	\prn{\frac{T_\text{R}}{10^{10}\,\text{GeV}}}^\frac{4}{5}.
\end{align}
At that time, the amplitude of Higgs is at most 
$h (t_\text{th}) \sim 5 \times 10^{10}\, \text{GeV}\, (0.01/\lambda)^{1/4}$ owing to the cosmic expansion,
even if we neglect the thermal dissipation of Higgs for $h > T$~\cite{Mukaida:2012qn}.
Once the amplitude of Higgs becomes smaller than the cosmic temperature $T$,
the dissipation rate of Higgs may be estimated as $\Gamma_{\rm dis}\sim \alpha^2 T$.
Then, the dissipation rate of Higgs at $t_\text{th}$ is simply given by $\Gamma_\text{dis} \sim \alpha^2 T_\text{th}$,
which is larger than the Hubble parameter at $t_\text{th}$.
Therefore, we expect that the whole system including Higgs is thermalized 
at least by $t_\text{th}$.

Now we conservatively impose the condition on the moduli mass as $m_\phi \lesssim C_H\mathcal H_{T=T_{\rm th}}$, \textit{i.e.},
\begin{align}
	m_\phi \lesssim 
	3 \times 10^{9} \, \text{GeV}\, C_H
	\prn{ \frac{\alpha}{0.1} }^\frac{16}{5} \prn{ \frac{10^{13}\,\text{GeV}}{m_I} }^\frac{4}{5} 
	\prn{ \frac{T_\text{R}}{10^{10}\, \text{GeV}} }^\frac{6}{5}.
	\label{m_up}
\end{align}
It ensures that the Higgs is thermalized before the moduli starts to oscillate.
Once thermalized, we can rely on the standard analysis based on the thermal bounce calculation.
Thus, the vacuum metastability is maintained for the center value of 
the Higgs and top quark masses even after the moduli oscillation~\cite{Rose:2015lna}.
(For some subtle issues relating to the moduli oscillation, see the next subsection.)

We derived a conservative upper bound on the moduli mass (\ref{m_up}), but there is also a lower bound.
This is because the moduli we have introduced couples to SM particles via Planck-suppressed interactions,
and hence it generally causes a cosmological moduli problem~\cite{Coughlan:1983ci,Ellis:1986zt,Goncharov:1984qm,Banks:1993en,deCarlos:1993wie}.
In order for the moduli to decay well before the big-bang nucleosynthesis begins, we need $m_\phi \gtrsim 100$\,TeV.
Anyway, there is a window for the moduli mass for the present solution to the vacuum stability problem to work.

\subsection{After moduli oscillation}

\begin{figure}[t]
\begin{center}
\begin{tabular}{cc}
\includegraphics[width=\linewidth]{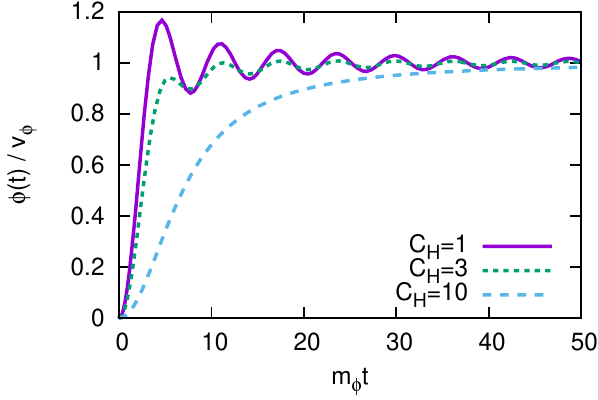}
\end{tabular}
\end{center}
\caption {\small
	Time evolution of the moduli $\phi$ for $C_H=1, 3$ and $10$.
}
\label{fig:phi}
\end{figure}

Now let us see more closely what happens after moduli 
starts to oscillate around its potential minimum: $\mathcal H \lesssim m_\phi $.
Fig.~\ref{fig:phi} shows the time evolution of the moduli $\phi$ for $C_H=1, 3$ and $10$.
We assume $v_\phi \lesssim M_P/C_H$ so that the moduli does not dominate the universe before the oscillation.

For $C_H=1$, the moduli reaches its maximum value $\phi_{\rm max}/v_\phi \simeq 1.17$ during the first oscillation.
It means that the effective quartic coupling at $\phi=\phi_{\rm max}$ takes
\begin{align}
	\lambda(\phi_{\rm max}) \simeq \lambda_0+0.17\xi_\lambda.
\end{align}
It is slightly (negatively) larger than the present quartic coupling at high energy scales
and hence it tends to make the vacuum unstable compared with the pure SM calculation.
Although it is difficult and beyond the scope of this letter to precisely analyze the vacuum stability in the presence of oscillating moduli or the rapidly changing Higgs potential, 
we may regard the time-dependent Higgs potential as if it is static as long as the temperature is much higher than the
moduli oscillation frequency.
Actually the last condition is satisfied for the most cases of our interest.
If we take $\xi_\lambda \simeq -0.02$ for the absolute stability during inflation, the relative enhancement of the
quartic coupling is about $0.3$ at the Planck scale, which does not significantly affect the vacuum stability
in thermal environment~\cite{Rose:2015lna}.

On the other hand, for $C_H\gtrsim 3$, the adiabatic suppression mechanism works~\cite{Linde:1996cx} and
the oscillation amplitude of the moduli is greatly reduced as seen in Fig.~\ref{fig:phi}.\footnote{
	The oscillation amplitude is in general {\it not} exponentially suppressed as $\sim \exp(-C_H)$ for large $C_H$ 
	as proved in Ref.~\cite{Nakayama:2011wqa}, but still the amplitude is suppressed enough to be neglected in our purpose.
}
In such a case, there is no enhancement of the Higgs quartic and top yukawa couplings 
since the moduli adiabatically relaxes to the potential minimum
and hence the vacuum stability issue reduces to the standard analysis based on the pure SM couplings.

\section{Discussion}  \label{sec:dis}

We have shown that the moduli-dependent Higgs quartic  (as well as other) couplings can make the
electroweak vacuum stable in the early universe.
This provides a simple solution to the vacuum stability problem during/after inflation.
We have demonstrated that
the stability of our vacuum is ensured with a natural choice of parameters; 
$-\xi_\lambda \sim \mathcal O (0.1)$ and $C_H \sim \mathcal O (1)$.

Our key idea is that the coupling constants may depend on the field value of some scalar fields.
Although we concentrated on the moduli-like field, there are lots of other possibilities.
For example, in models with abelian flavor symmetry~\cite{Froggatt:1978nt},
yukawa couplings are determined by the so-called flavon field~(for a recent brief review, 
see \textit{e.g.} Ref.~\cite{Bauer:2016rxs} and references therein) and it may be natural to consider the varying yukawa couplings along with the flavon dynamics.\footnote{
	Recently it is pointed out that the change of yukawa coupling in this class of models
	may lead to the first-order electroweak phase transition~\cite{Baldes:2016rqn}.
}
In this class of models, a complex scalar $S$, charged under U(1)$_F$, couples to SM fields like
\begin{align}
	\mathcal L = \left( \frac{S}{M}\right)^{n^u_{ij}} y^{u}_{ij} \overline Q_i \tilde H u_{R_j}
			+  \left( \frac{S}{M}\right)^{n^d_{ij}} y^{d}_{ij} \overline Q_i H d_{R_j}  + {\rm h.c.},
\end{align}
where $M$ is a cutoff scale. Depending on the charge assignments of SM quarks under U(1)$_F$, the exponents $n_{ij}$ are fixed.
For $\langle S\rangle/M \sim 0.23$ and appropriate charge assignments, all the yukawa couplings $y_{ij}$ can be $\mathcal O(1)$
and hence the hierarchy of quark masses as well as quark mixing angles are naturally explained.
Unfortunately, the top yukawa entry has $n^u_{33}=0$ since the top yukawa coupling is already 
$\mathcal O(1)$ without introducing flavon in a typical flavon model .
However, it is allowed to add a flavon-dependent term
\begin{align}
	\mathcal L \simeq -\frac{|S|^2}{M^2}|H|^4 
	+\left(\frac{\left |S\right|^2}{M^2} y^{u}_{33} \overline Q_3 \tilde H t_{R} + {\rm h.c.}\right),
\end{align}
without conflicting the symmetry.
This can modify the Higgs quartic and top yukawa couplings up to $(\langle S\rangle /M)^2 \sim 0.05$ depending on the flavon dynamics in the early universe.
As we have seen, an $\mathcal{O}(0.01)$ change in the quartic and/or top yukawa 
is enough to make sure that the vacuum is absolutely stable.
Note that the cosmology is somewhat nontrivial because the spontaneous breaking of U(1)$_F$ symmetry after inflation leads 
to cosmic strings, and if there is a (small) explicit breaking term,
stable or unstable domain walls will appear depending on the breaking pattern
(See \textit{e.g.}, Refs.~\cite{Preskill:1991kd,Riva:2010jm} for related issues).
We leave these issues for future work.

\section*{Acknowledgments}
\small

This work was supported by the Grant-in-Aid for Scientific Research on Scientific Research A (No.26247042 [KN]),
Young Scientists B (No.26800121 [KN]) and Innovative Areas (No.26104009 [KN], No.15H05888 [KN]),
World Premier International Research Center Initiative (WPI Initiative), MEXT, Japan (K.M.\ and K.N.),
JSPS Research Fellowships for Young Scientists (Y.E.\ and K.M.),
and the Program for Leading Graduate Schools, MEXT, Japan (Y.E.).


\bibliographystyle{apsrev4-1}
\bibliography{ref}
\end{document}